\documentclass{article} 
\usepackage{iclr2026_conference,times}

\usepackage{amsmath,amsfonts,bm}

\def\eqref#1{equation~\ref{#1}}

\def\1{\bm{1}}

\DeclareMathAlphabet{\mathsfit}{\encodingdefault}{\sfdefault}{m}{sl}
\SetMathAlphabet{\mathsfit}{bold}{\encodingdefault}{\sfdefault}{bx}{n}

\usepackage{graphicx,booktabs}
\usepackage{hyperref}
\usepackage{url}

\title{Inference-Time Safety for Code LLMs via Retrieval-Augmented Revision}

\author{
Manisha Mukherjee \\
Carnegie Mellon University \\
Pittsburgh, PA, USA \\
\texttt{mmukherj@andrew.cmu.edu}
\And
Vincent J. Hellendoorn \\
Carnegie Mellon University \\
Pittsburgh, PA, USA \\
\texttt{vhellendoorn@cmu.edu}
}

\iclrfinalcopy 
\begin{document}

\maketitle

\begin{abstract}
Large Language Models (LLMs) are increasingly deployed for code generation in high-stakes software development, yet their limited transparency in security reasoning and brittleness to evolving vulnerability patterns raise critical trustworthiness concerns. Models trained on static datasets cannot readily adapt to newly discovered vulnerabilities or changing security standards without retraining, leading to the repeated generation of unsafe code.

We present a principled approach to trustworthy code generation by design that operates as an inference-time safety mechanism. Our approach employs retrieval-augmented generation to surface relevant security risks in generated code and retrieve related security discussions from a curated Stack Overflow knowledge base, which are then used to guide an LLM during code revision. This design emphasizes three aspects relevant to trustworthiness: (1) interpretability, through transparent safety interventions grounded in expert community explanations; (2) robustness, by allowing adaptation to evolving security practices without model retraining; and (3) safety alignment, through real-time intervention before unsafe code reaches deployment.

Across real-world and benchmark datasets, our approach improves the security of LLM-generated code compared to prompting alone, while introducing no new vulnerabilities as measured by static analysis. These results suggest that principled, retrieval-augmented inference-time interventions can serve as a complementary mechanism for improving the safety of LLM-based code generation, and highlight the ongoing value of community knowledge in supporting trustworthy AI deployment.
\end{abstract}

\section{Introduction}
Large language models are widely deployed for automated code generation in modern software development workflows, underpinning a growing ecosystem of code assistants and integrated development tools, including GitHub Copilot, ChatGPT, Amazon Q, Cursor, and similar systems \cite{GitHubCopilot,ChatGPT,CodeWhisperer,Cursor}.
 These systems have demonstrated clear benefits for developer productivity, but their use in security-sensitive settings raises important trustworthiness concerns. LLMs trained on large collections of open-source code may inherit vulnerable or outdated coding patterns from their training data, including patterns associated with known Common Weakness Enumerations (CWEs) \cite{pearce2025asleep}. At the same time, programming languages, libraries, and frameworks evolve rapidly. For example, major platforms such as TensorFlow release new versions frequently, often deprecating unsafe APIs and patching previously unknown vulnerabilities \cite{TensorFlowReleases}. As a result, models trained on static snapshots of code repositories may continue to generate code that no longer reflects current security best practices.

These limitations are particularly concerning because developers often place substantial trust in LLM-generated code and may integrate it into production systems with limited security review \cite{kabir2024stack,jiao2025generative}. In adversarial settings, such trust can lead to exploitable vulnerabilities, data breaches, or degraded system reliability. While LLMs can produce syntactically correct and functionally valid code, they frequently lack transparent reasoning about security risks and may overlook subtle, context-dependent implications of specific libraries or usage patterns. Addressing these shortcomings through retraining or fine-tuning alone is costly and typically infrequent, limiting the ability of deployed models to adapt to newly discovered vulnerabilities or evolving security standards.

In contrast, developer communities such as Stack Overflow represent a continuously evolving source of security-relevant knowledge. Over more than a decade, Stack Overflow has enabled practitioners to identify, discuss, and revise insecure coding practices through questions, answers, and community comments. These comments often highlight security concerns, explain why particular approaches are risky, and suggest safer alternatives. At the same time, the quality of Stack Overflow content varies, and answers or code snippets may be incomplete, outdated, or context-dependent. Nonetheless, Stack Overflow discussions are subject to ongoing community review and revision, allowing problematic guidance to be flagged over time and providing human-authored explanations that reflect evolving security reasoning.

\begin{figure}[t]
    \centering
    \includegraphics[width=0.7\columnwidth]{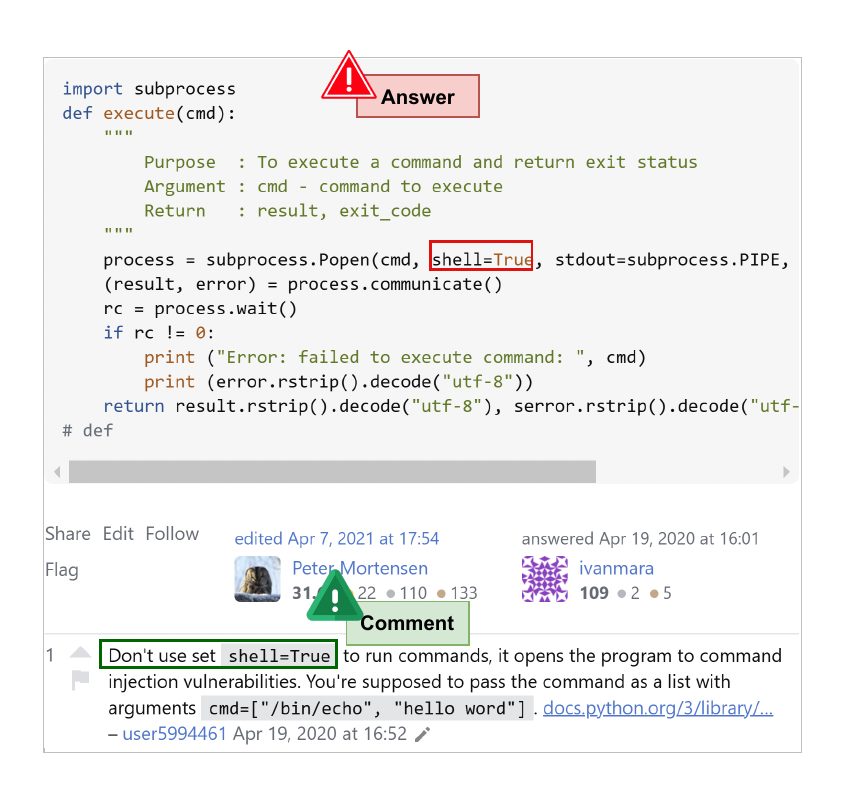}
  \vspace{-5mm}
    \caption{AnswerID: 61307412, which includes community comments providing security insights. This content was used as context to enhance the generated code in SOSecure.}

    \label{fig:hero}
\end{figure}

Motivated by this contrast, we investigate whether community knowledge can be leveraged as an inference-time safety mechanism for LLM-based code generation. Prior work has explored retrieval-augmented generation to incorporate external knowledge from sources such as documentation or vulnerability databases, typically to inform generation at prompt time \cite{lewis2020retrieval,vulrag}. In contrast, we focus on Stack Overflow as a distinct knowledge source that reflects the evolution of human security understanding over time, where vulnerabilities are identified, debated, and refined through community discussion. By retrieving these discussions after code has been generated, we explore whether community-authored security reasoning can serve as effective inference-time guidance for revising potentially unsafe outputs.

To this end, we introduce \textsc{SOSecure}, a retrieval-augmented approach that operates after code generation to revise potentially unsafe outputs. Given an LLM-generated code snippet, \textsc{SOSecure} retrieves vulnerability-oriented Stack Overflow answers and comments that discuss similar code patterns and explicitly mention security concerns. These discussions are then provided as contextual guidance to the LLM, which may revise the code accordingly or determine that no changes are necessary.

We position \textsc{SOSecure} as a complementary safety layer rather than 
a complete solution to secure code generation. By incorporating 
community-authored explanations at inference time, \textsc{SOSecure} is 
designed to support key principles of trustworthy AI, including 
interpretability through transparent safety interventions grounded in 
human reasoning, robustness to evolving security practices without model 
retraining, and safety alignment through real-time intervention before 
unsafe code reaches deployment. Through evaluation on multiple datasets of 
LLM-generated code, we show that this inference-time intervention improves 
security outcomes compared to prompting alone, while introducing no new 
vulnerabilities as measured by static analysis. Together, these results 
suggest that community-driven retrieval can serve as a principled 
component of inference-time safety for LLM-based code generation.

\section{Motivating Example}

\begin{figure*}[!t]
    \centering
    \includegraphics[width=1.03\columnwidth]{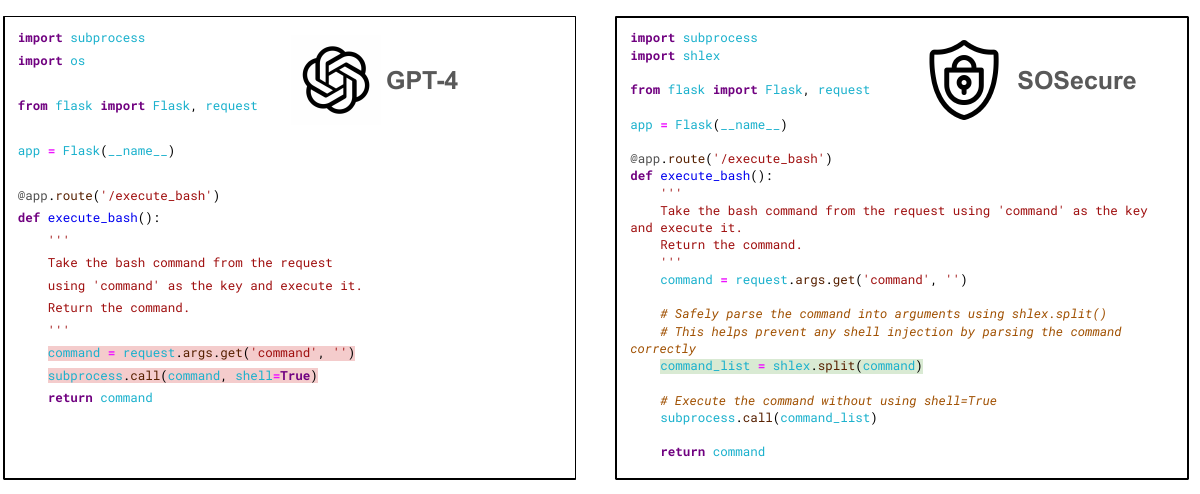}
  
    \vspace{-3mm}
    \caption{Inference-time revision guided by community knowledge. Left: original GPT-4
output containing a usage pattern commonly associated with CWE-078. Right:
one possible revised output produced after the model is provided with a
relevant Stack Overflow discussion (Figure~\ref{fig:hero}) as contextual
guidance.
}
    
 \vspace{-10pt}
 
    \label{fig:codes}
\end{figure*}

Figure~\ref{fig:codes} illustrates a motivating example of our approach. 
A user prompts an LLM to generate code for executing bash commands in a 
Flask application. The model produces functionally correct code that 
invokes \texttt{subprocess.call()} with \texttt{shell=True}, a usage 
pattern commonly associated with command injection risks (CWE-078).

When this output is processed by \textsc{SOSecure}, the system retrieves 
a Stack Overflow discussion in which community members caution against 
this pattern, noting that ``using \texttt{shell=True} opens the program 
to command injection vulnerabilities.'' This discussion is provided to 
the LLM as contextual guidance during inference-time revision. In 
response, the model produces an alternative implementation that avoids 
the risky usage pattern.

\section{Method: Inference-Time Safety via Community-Driven Retrieval}

We present \textsc{SOSecure}, an inference-time safety mechanism designed to
improve the security of LLM-generated code by incorporating community-authored
security knowledge during post-generation revision. The core idea is to surface
relevant security discussions from Stack Overflow after an LLM has produced
code, and to use these discussions as contextual guidance when the model assesses
whether revisions are warranted. \textsc{SOSecure} is model-agnostic, requires no
retraining or fine-tuning, and is intended to complement existing training-time
and analysis-based security techniques.

\subsection{Design Overview}

\begin{figure*}[!t]
    \centering    
     \vspace{-7mm}
\includegraphics[width=1\linewidth]{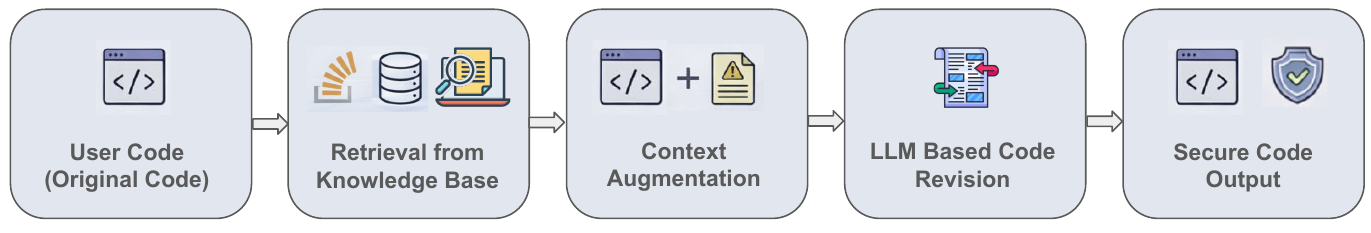}

    \caption{Conceptual overview of inference-time revision with \textsc{SOSecure}. Community security
discussions from Stack Overflow are retrieved based on similarity to
LLM-generated code and provided as contextual guidance during revision.
}

    \label{fig:framework}
\end{figure*}

Figure~\ref{fig:framework} provides a high-level overview of the \textsc{SOSecure}
workflow. Given an initial code snippet generated by an LLM, the system performs
three steps: (1) retrieval of security-relevant community discussions that address
similar code patterns, (2) construction of a revision prompt that presents these
discussions as advisory context, and (3) inference-time revision in which the LLM
may update the code or retain the original output. Retrieved content is not treated
as ground truth and is never directly injected into the code. Instead, the model is
prompted to reason about potential security implications in light of community
feedback.

\subsection{Security-Oriented Knowledge Base}
\label{sec:knowledge-base}

To support inference-time retrieval, we construct a security-oriented knowledge base from Stack Overflow. We focus on answers and their associated comment threads that explicitly reference security concerns, vulnerabilities, or unsafe practices. Following prior work on security content in developer communities \cite{meng2018secure,rahman2019snakes,hong2021dicos}, we filter posts using a curated set of security-related keywords, including references to known weaknesses, deprecated functionality, and risky usage patterns.

As a minimal quality control step, we require that either the answer or at least one associated comment has received at least one community upvote. This criterion removes clearly unendorsed content while remaining intentionally lightweight. We do not assume that upvotes or other community metadata provide guarantees of correctness or security, particularly given the context-dependent and evolving nature of security best practices. Instead, this design prioritizes recall over precision during retrieval, under the principle that the downstream LLM is better suited than static filters to reason about the relevance and validity of surfaced security critiques.

Accordingly, retrieved Stack Overflow discussions are treated as auxiliary signals that surface potential risks, alternatives, or points of concern, rather than as authoritative prescriptions. These human-authored explanations are provided as contextual guidance at inference time, allowing the language model to decide whether and how to revise the generated code.

\subsection{Retrieval of Community Discussions}

Given an LLM-generated code snippet, \textsc{SOSecure} retrieves relevant entries
from the knowledge base based on lexical similarity between the generated code and
Stack Overflow code snippets. We employ a BM25-based retrieval model, which is well
suited for code- and API-centric queries where identifier overlap carries important
semantic information. Prior work has shown that lexical retrieval methods such as
BM25 remain highly competitive with, and in some cases outperform, more complex
neural retrievers in sparse, keyword-driven settings \cite{guo2020deep}. Related
work in code analysis has likewise observed that lexical similarity is often a
strong signal for identifying related code patterns and usage contexts
\cite{chakraborty2021deep}.

In preliminary experiments, we compared dense embedding-based retrieval with sparse lexical retrieval and found BM25 to be more reliable for surfacing security-relevant discussions tied to specific libraries or usage patterns. Dense methods frequently failed to retrieve relevant comments when vulnerabilities hinged on concrete API calls, configuration flags, or error messages. This behavior is consistent with observations that dense embeddings may underweight rare tokens and syntactic cues that are critical for reasoning about code security.

In contrast, BM25 more consistently matched generated code to Stack Overflow
discussions that explicitly reference the same functions, parameters, or usage
patterns (e.g., \texttt{shell=True}, \texttt{pickle.loads}, or \texttt{debug=True}). Since many community security warnings are triggered by such concrete identifiers rather than high-level semantic similarity, we adopt BM25 as a principled design choice that prioritizes precision on security-critical cues. For each input snippet, the system retrieves the top-$k$ most similar Stack Overflow answers along with their associated comment threads. Unless otherwise specified, we use $k=5$, which provided a balance between contextual richness and prompt length in our experiments.

\subsection{Inference-Time Revision Prompt}

Retrieved discussions are incorporated into a structured revision prompt alongside
the original code. The prompt instructs the LLM to review its generated code in light
of the provided community feedback and to determine whether any security-relevant
changes are appropriate. The model is explicitly permitted to leave the code
unchanged if it determines that the original implementation already follows secure
practices.

This conservative prompting strategy is intentional. By framing retrieved content
as contextual guidance rather than prescriptive instructions, \textsc{SOSecure}
avoids enforcing potentially outdated or incomplete advice and instead relies on
the model's reasoning capabilities to weigh community input against the code at
hand. This design supports interpretability by making the basis for revision
transparent and reduces the risk of introducing new vulnerabilities through
overcorrection.

\section{Background}

\paragraph{Security in LLM-Generated Code.}
Large language models specialized for code generation 
\cite{xu2022systematic,li2023starcoder,fried2022incoder,austin2021program} 
can produce syntactically correct and functionally valid solutions, but 
frequently generate code with security vulnerabilities. Prior studies have 
shown that instruction-tuned models such as ChatGPT produce insecure code 
in many security-sensitive scenarios 
\cite{khoury2023secure}. While some approaches seek to improve security 
through fine-tuning or specialized training 
\cite{sven,wang2023enhancing,sallm,franc,chakraborty2021deep}, ensuring 
secure behavior in LLM-generated code remains an open challenge, 
particularly in settings where models must adapt to evolving 
vulnerabilities without retraining.

\paragraph{Vulnerability Detection and Evaluation Datasets.}
We evaluate security using static analysis tools including CodeQL 
\cite{codeql}, an industry-standard analyzer supporting multiple 
programming languages, and Bandit \cite{bandit}, a Python-specific 
security linter. Both tools have been widely used to evaluate the 
security of LLM-generated code and provide a practical, automated means 
of identifying common vulnerabilities \cite{sallm,franc}. Our evaluation 
spans three complementary datasets: SALLM \cite{sallm}, which focuses on 
code-format prompts; LLMSecEval \cite{llmseceval}, which uses natural 
language prompts; and LMSys \cite{lmsys}, which captures real-world user 
conversations with LLMs. Together, these datasets address known challenges 
in constructing and evaluating vulnerability benchmarks for generative 
models \cite{peng2025cweval}.

\paragraph{Security Knowledge in Stack Overflow.}
Prior work has examined the role of developer communities in shaping 
secure coding practices 
\cite{mukherjee2023stack,meyers2019pragmatic,le2021large}. Empirical 
studies have shown that Stack Overflow contains a mixture of secure 
and insecure examples, reflecting the diversity and evolving nature of 
community-contributed content \cite{meng2018secure,rahman2019snakes}. \cite{hong2021dicos} demonstrated that community comments can 
signal insecure code by highlighting risks and contextual concerns. We 
build on this line of work by explicitly filtering for security-oriented 
Stack Overflow discussions and using community-authored explanations as contextual 
guidance during inference-time revision of LLM-generated code.

\paragraph{Retrieval-Augmented Generation for Security.}
Retrieval-augmented generation (RAG) \cite{lewis2020retrieval} enables 
models to incorporate external knowledge at inference time without 
retraining and has been applied to security-related tasks such as 
vulnerability detection using CVE instances \cite{vulrag}. In this work, 
we adopt RAG as an inference-time safety mechanism for code generation, 
retrieving community-identified security discussions from Stack Overflow 
after code has been generated. This post-generation design reflects a 
safety-oriented approach that allows external knowledge to inform 
revision of potentially unsafe outputs while remaining complementary to 
training-time methods and static analysis techniques.

\begin{table}[!t]
\centering
\small
\begin{tabular}{lccccc}
\toprule
\textbf{Dataset} & \textbf{Prompt-only} & \textbf{GPT-4+CWE} & \textbf{\textsc{SOSecure}} & $\Delta$ \textbf{Fix Rate} & \textbf{Intro Rate (SOSecure)} \\
\midrule
SALLM           & 49.1\% & 58.5\% & 71.7\% & +22.6\% & 0.0\% \\
LLMSecEval  & 56.5\% & 69.6\% & 91.3\% & +34.8\% & 0.0\% \\
LMSys           & 37.5\% & 45.8\% & 96.7\% & +59.2\% & 0.0\% \\
\bottomrule
\end{tabular}
\caption{Security outcomes across three datasets. \textbf{Fix Rate} measures the
fraction of previously flagged vulnerabilities no longer detected after
revision. \textbf{GPT-4+CWE} provides the model with an explicit vulnerability
label at inference time. \textbf{Intro Rate (SOSecure)} measures the fraction of
samples for which new vulnerabilities are introduced.}
\vspace{-10pt}
\label{tab:main-results}
\end{table}

\section{Evaluation}

We evaluate whether inference-time community-driven retrieval influences the
security of LLM-generated code and assess the extent to which it improves
security outcomes without introducing new vulnerabilities. Our evaluation is
designed to isolate the effect of inference-time retrieval rather than to
benchmark state-of-the-art vulnerability repair systems.

\subsection{Experimental Setup}

\paragraph{Datasets.}
We evaluate \textsc{SOSecure} on three complementary datasets that capture
different prompting styles and usage contexts.

\textbf{SALLM}~\cite{sallm} contains 100 security-oriented prompts provided in
both natural language and code formats, with corresponding LLM-generated code
covering 45 vulnerability types (CWEs). Each sample is mapped to a specific CWE.
We retain only samples associated with CWEs supported by default CodeQL queries,
resulting in 74 samples used for evaluation.

\textbf{LLMSecEval}~\cite{llmseceval} is a natural language prompt-to-code dataset
derived from Pearce et al.~\cite{pearce2025asleep}. It consists of 150 prompts,
including 83 Python and 67 C code generations. We similarly restrict evaluation
to samples flagged under CWEs supported by default CodeQL queries, yielding 49
Python and 40 C samples.

\textbf{LMSys}~\cite{lmsys} reflects real-world deployment scenarios, drawing from
the LMSys-Chat-1M corpus of user interactions with LLMs. Of the one million
conversations, 43,269 contain Python code, and 31,008 consist of single-round
conversations with a single code block. To construct a high-quality subset
containing genuine vulnerabilities, we apply a two-stage filtration process.
First, we retain only samples flagged as vulnerable by both Bandit~\cite{bandit}
and CodeQL~\cite{codeql}, resulting in 2,809 samples. We then restrict to CWEs
supported by default CodeQL queries, yielding a final set of 240 Python samples
used for analysis.

Across all datasets, we focus exclusively on LLM-generated code that is flagged
as potentially vulnerable by static analysis tools, following prior work
\cite{sallm,franc}. This allows us to directly evaluate whether inference-time
revision reduces observed security issues.

\paragraph{Models and Baselines.}
We use GPT-4 as the underlying code generation model across all experiments.
We compare \textsc{SOSecure} against three prompt-based baselines that reflect
common and increasingly sophisticated usage patterns in practice.
\emph{Prompt-only} corresponds to standard code generation without any
additional security guidance.
\emph{Revision-only} prompts the model to revise its own output for security
issues without any retrieved context.
Finally, \emph{GPT-4+CWE} provides the model with an explicit vulnerability
label (e.g., CWE identifier) at inference time, representing an oracle-style
baseline that assumes prior identification of the weakness but does not supply
explanatory or contextual security knowledge.
This baseline allows us to isolate the effect of community-authored security
discussions beyond simply naming the vulnerability.

\paragraph{Metrics.}
We assess security outcomes using CodeQL \cite{codeql} and Bandit \cite{bandit}.
We report \textbf{Fix Rate} (FR), the fraction of previously flagged
vulnerabilities no longer detected after revision, and \textbf{Introduction
Rate} (IR), the fraction of samples for which new vulnerabilities are
introduced.

\subsection{Main Results}

Table~\ref{tab:main-results} summarizes security outcomes across all three
datasets. \textsc{SOSecure} consistently improves fix rates compared to
prompt-only generation, with gains ranging from 22.6 percentage points on
SALLM to 59.2 percentage points on LMSys. Across all datasets, \textsc{SOSecure} outperforms GPT-4+CWE, indicating that
explicit vulnerability labels alone are insufficient and that community
explanations provide additional, actionable security guidance at inference time.
Across all datasets, no new
vulnerabilities are introduced as measured by static analysis.

These results indicate that inference-time community-driven retrieval can
substantially influence model behavior while avoiding systematic
overcorrection.

\subsection{Ablation: Effect of Community-Driven Retrieval}

Table~\ref{tab:ablation} isolates the contribution of retrieval on the LMSys
dataset. Prompting the model to revise its own output yields only marginal
improvements over prompt-only generation. In contrast, adding retrieved
community discussions leads to a large increase in fix rate, highlighting the
central role of community-driven context in inference-time safety decisions.

\subsection{Evaluation on C Code}

While most evaluated samples are written in Python, we additionally assess
\textsc{SOSecure} on C code from the LLMSecEval dataset to examine behavior
beyond a single programming language. Table~\ref{tab:c_results} reports system-level metrics on C code.

On C code, \textsc{SOSecure} achieves a Fix Rate of 73.3\%, compared to 53.3\%
for GPT-4 and 60.0\% for GPT-4+CWE, while introducing no new vulnerabilities.
Analysis by CWE type shows strong performance on several high-severity
vulnerabilities, including CWE-078 (OS Command Injection) and CWE-190 (Integer
Overflow), while highlighting limitations on others such as CWE-022 (Path
Traversal). These results suggest that community-driven inference-time guidance
may generalize across programming languages supported by static analysis, without
language-specific tuning.

\begin{table}[t]
\centering
\small
\begin{tabular}{lcc}
\toprule
\textbf{Method} & \textbf{Fix Rate} & \textbf{Intro Rate} \\
\midrule
Prompt-only baseline              & 37.5\% & 0.0\% \\
GPT-4+CWE (label-only prompting)  & 45.8\% & 0.0\% \\
Revision-only (no retrieval)      & 41.2\% & 0.0\% \\
\textsc{SOSecure} ($k=5$)         & 96.7\% & 0.0\% \\
\bottomrule
\end{tabular}
\caption{Ablation on LMSys isolating the effect of community-driven retrieval.
Providing explicit vulnerability labels (GPT-4+CWE) or prompting the model to
revise its own output yields modest improvements, while incorporating retrieved
community discussions leads to substantially higher fix rates.}
\label{tab:ablation}
\end{table}

\section{Discussion}

Our results suggest that inference-time community-driven retrieval can serve as
an effective and lightweight mechanism for improving the security of
LLM-generated code. Rather than relying on retraining or vulnerability-specific
repair models, \textsc{SOSecure} leverages human-authored security reasoning
embedded in developer discussions to influence model behavior at the point of
generation. This design allows security-relevant context to be introduced
precisely when it is needed, even when vulnerabilities are implicit or not
explicitly requested by the user.

A key observation from our evaluation is that self-revision alone yields only
marginal improvements, while the addition of retrieved community discussions
leads to substantial gains. This suggests that the benefits of
\textsc{SOSecure} stem not from prompting the model to reconsider its output,
but from exposing it to concrete explanations of why certain patterns are
unsafe. These explanations often provide causal insight, contextual constraints,
or practical guidance that is difficult to encode in static prompts or
fine-tuned parameters. In this sense, community-driven retrieval helps bridge
the gap between memorized knowledge and its appropriate application in
security-sensitive contexts.

Notably, even when provided with explicit CWE labels, GPT-4 benefits from
community-authored explanations, suggesting that trustworthiness improvements
require more than vulnerability identification alone.

We view \textsc{SOSecure} as a complementary safety layer rather
than a replacement for existing approaches. Static analysis, secure coding
guidelines, and training-time interventions all play important roles in
improving software security. Inference-time retrieval offers a different point
in the design space, enabling deployed models to incorporate evolving security
knowledge without retraining. This makes the approach particularly suitable for
settings where models are already deployed and must adapt to changing security
practices over time.

\begin{table}[t]
\centering
\small
\begin{tabular}{lccc}
\toprule
\textbf{Method} & \textbf{Fix Rate} & \textbf{Intro Rate} & \textbf{No Change Rate} \\
\midrule
Prompt-only            & 53.3\% & 0.0\% & 80.0\% \\
GPT-4+CWE              & 60.0\% & 0.0\% & 77.5\% \\
\textsc{SOSecure}      & 73.3\% & 0.0\% & 72.5\% \\
\bottomrule
\end{tabular}
\caption{Performance on C code from the LLMSecEval dataset.
\textbf{Fix Rate} measures the fraction of previously flagged vulnerabilities
that are no longer detected after revision.
\textbf{Intro Rate} captures newly introduced vulnerabilities.
\textbf{No Change Rate} denotes cases where the model produces identical code
after the revision prompt.}
\label{tab:c_results}
\end{table}

\paragraph{Illustrative Example.}

Beyond direct pattern avoidance, SOSecure can leverage community discussions
to support more indirect security reasoning. One representative example
involves unsafe serialization in a web application, where user input was
stored and retrieved using Python’s \texttt{pickle} module. Although the
original code did not rely on \texttt{eval}, it was flagged by static analysis
due to the risks of deserializing untrusted data.

The retrieved Stack Overflow discussion did not explicitly warn against
\texttt{pickle}, but emphasized that dynamic code execution mechanisms are
dangerous in web-facing contexts and should be avoided in favor of safer data
formats. Guided by this contextual explanation, SOSecure revised the code to
use JSON-based serialization instead, eliminating the deserialization risk
while preserving the original program behavior.

Notably, when prompted to revise the same code without retrieved community
context, GPT-4 did not apply this change. This example illustrates how
community-authored explanations can support inference-time security reasoning
that goes beyond direct pattern matching or keyword-based fixes.

\section{Implications for Trustworthy AI Design}

Our findings suggest several broader implications for the design of trustworthy
AI systems for code generation. Rather than treating trustworthiness solely as a
property to be enforced at training time, this work highlights the value of
\emph{inference-time interventions} that can adapt model behavior to evolving
knowledge and deployment contexts. In settings where retraining is costly or
infrequent, inference-time mechanisms offer a practical means of improving
system behavior after deployment.

A central design insight from \textsc{SOSecure} is the use of human-authored
explanations as a safety signal. Community discussions on Stack Overflow often
capture not only what code is insecure, but why it is insecure, including
contextual constraints, common misconceptions, and trade-offs. By surfacing
these explanations during inference, the system provides a more transparent and
interpretable form of safety intervention than opaque model updates or static
filters. This suggests that incorporating structured human reasoning can be a
useful complement to automated detection and enforcement mechanisms.

The approach also illustrates how trustworthy behavior can emerge from
\emph{compositional system design}. \textsc{SOSecure} does not modify the
underlying model or replace existing security tools; instead, it composes
retrieval, prompting, and static analysis into a lightweight safety layer. This
modularity allows the system to evolve alongside external knowledge sources and
to integrate with other safeguards, such as static analyzers or human review,
without requiring tight coupling or retraining.

Finally, our results underscore the continuing relevance of developer
communities in the AI-assisted programming ecosystem. While large language
models increasingly mediate access to programming knowledge, community forums
remain an active venue for identifying emerging vulnerabilities and correcting
unsafe practices. Designing AI systems that can leverage, rather than replace,
these collective processes may be an important step toward maintaining both
technical robustness and social accountability in AI-supported software
development.

\section{Limitations}

Our work has several limitations. First, our evaluation relies on static
analysis tools, which are known to exhibit false positives and false negatives.
While we follow established practice and use widely adopted tools, static
analysis does not capture all security properties, and our results should be
interpreted accordingly.

Second, \textsc{SOSecure} depends on the quality and relevance of retrieved
Stack Overflow discussions. While community review helps surface security
concerns over time, Stack Overflow content can be incomplete, outdated, or
context-dependent. Although our filtering strategy reduces exposure to noisy
content, retrieval errors may still limit effectiveness in some cases.

Third, our retrieval mechanism is primarily lexical and does not explicitly
model semantic or structural code properties. As a result, retrieved discussions
may be only loosely related to the generated code, particularly for complex or
non-local vulnerabilities. Incorporating richer retrieval signals is a natural
direction for future work.

\paragraph{Illustrative Example}
We observe failure cases in which retrieved community discussions are outdated or
do not fully reflect current security best practices. For example, in some
SSL/TLS configuration scenarios, discussions referenced protocol versions that
were considered acceptable at the time but are now deprecated. In such cases,
\textsc{SOSecure} partially addressed flagged issues without fully updating the
configuration, highlighting a limitation of relying on evolving community
knowledge for security guidance.

Finally, our evaluation focuses on security outcomes and does not include systematic functional testing, as most datasets lack executable test cases. To guard against trivial regressions, we manually inspected a random subset of revised outputs (10 samples per dataset) and verified that the produced code remained syntactically valid and consistent with the original intent. While we did not observe cases where SOSecure broke functionality in these checks, this process does not provide formal guarantees. Incorporating automated testing, program
analysis, or differential execution techniques to more rigorously assess
functional correctness is a direction for future work.

\section{Conclusion}

We presented \textsc{SOSecure}, an inference-time approach for improving the
security of LLM-generated code by leveraging community-driven security
discussions from Stack Overflow. By introducing retrieved human-authored
explanations during code revision, the approach provides a lightweight mechanism
for influencing model behavior without retraining or language-specific tuning.
Our evaluation across multiple datasets and programming languages shows that
community-driven retrieval can substantially improve security outcomes compared
to prompting alone, while avoiding the introduction of new vulnerabilities.
Rather than replacing existing security techniques, inference-time retrieval
offers a complementary design strategy that enables deployed models to adapt to
evolving security practices. We hope this work encourages further exploration
of principled inference-time interventions as a component of trustworthy AI
systems for code generation.

\section{Data Availability}

All code, datasets, prompts, and evaluation artifacts are publicly available at 
\href{https://github.com/manishamukherjee/SOSecure}{https://github.com/manishamukherjee/SOSecure}.

\bibliography{iclr2026_conference}
\bibliographystyle{iclr2026_conference}


\end{document}